\documentclass[fleqn,usenatbib]{mnras}

\usepackage{newtxtext,newtxmath}

\usepackage[T1]{fontenc}

\DeclareRobustCommand{\VAN}[3]{#2}
\let\VANthebibliography\thebibliography
\def\thebibliography{\DeclareRobustCommand{\VAN}[3]{##3}\VANthebibliography}

\usepackage{graphicx}	
\usepackage{amsmath}	
\usepackage{todonotes}

\title[WASP-186,187]{WASP-186 and WASP-187: two hot Jupiters discovered by SuperWASP and SOPHIE with additional observations by TESS}

\author[N. Schanche et al.]{
N. Schanche,$^{1}$\thanks{E-mail: ns81@st-andrews.ac.uk}
G. Hébrard,$^{2,3}$
A. Collier Cameron,$^{1}$
S. Dalal,$^{2}$ 
B. Smalley,$^{4}$\newauthor
T. G. Wilson,$^{1}$
I. Boisse,$^{5}$
F. Bouchy,$^{6}$
D.J.A Brown,$^{7,8}$
O. Demangeon,$^{9}$\newauthor
C.A. Haswell,$^{10}$ 
C. Hellier,$^{4}$
U.C. Kolb,$^{10}$
T. Lopez,$^{5}$
P.F.L. Maxted,$^{4}$\newauthor
D.L. Pollacco,$^{7,8}$
R.G. West,$^{7,8}$
P.J. Wheatley$^{7,8}$
\\
\\
$^{1}$Centre for Exoplanet Science, SUPA, School of Physics and Astronomy, University of St Andrews, St Andrews KY16 9SS, UK\\
$^{2}$Institut d'astrophysique de Paris, UMR7095 CNRS, Universit\'e Pierre \& Marie Curie, 98bis boulevard Arago, 75014 Paris, France\\
$^{3}$Observatoire de Haute-Provence, CNRS, Universit\'e d'Aix-Marseille, 04870 Saint-Michel-l'Observatoire, France\\
$^{4}$Astrophysics Group, Keele University, Staffordshire, ST5 5BG, UK \\
$^{5}$Laboratoire d'Astrophysique de Marseille, Universit\'e de Provence, UMR6110 CNRS, 38 rue F. Joliot Curie, 13388  Marseille cedex 13, France \\
$^{6}$Observatoire de Gen\`eve,  Universit\'e de Gen\`eve, 51 Chemin des Maillettes, 1290 Sauverny, Switzerland \\
$^{7}$Department of Physics, University of Warwick, Gibbet Hill Road, Coventry, CV4 7AL, UK \\
$^{8}$Centre for Exoplanets and Habitability, University of Warwick, Gibbet Hill Road, Coventry, CV4 7AL, UK \\
$^{9}$Instituto de Astrof{\'\i}sica e Ci\^encias do Espa\c{c}o, Universidade do Porto, CAUP, Rua das Estrelas, 4150-762 Porto, Portugal \\
$^{10}$School of Physical Sciences, The Open University, Milton Keynes, MK7 6AA, UK \\
}

\date{Accepted XXX. Received YYY; in original form ZZZ}

\pubyear{2020}

\begin{document}
\label{firstpage}
\pagerange{\pageref{firstpage}--\pageref{lastpage}}
\maketitle

\begin{abstract}
We present the discovery of two new hot Jupiters identified from the WASP survey, WASP-186b and WASP-187b (TOI-1494.01 and TOI-1493.01). Their planetary nature was established from SOPHIE spectroscopic observations, and additional photometry was obtained from TESS. Stellar parameters for the host stars are derived from spectral line, IRFM, and isochrone placement analyses. These parameters are combined with the photometric and radial velocity data in an MCMC method to determine the planetary properties. WASP-186b is a massive Jupiter (4.22$\pm{0.18}\ M_J$, 1.11 $\pm{0.03}\ R_J$) orbiting a mid-F star on a 5.03 day eccentric (e=0.327$\pm{0.008}$) orbit. WASP-187b is a low density (0.80 $\pm{0.09}\ M_J$, 1.64 $\pm{0.05} R_J$) planet in a 5.15 day circular orbit around a slightly evolved early F-type star. 
\end{abstract}




\section{Introduction}


While hot Jupiters are relatively rare, occurring for less than 1 star in 100 \citep{Zhou2019}, they are valuable because they are extreme examples of planetary-system formation. Because these planets have deep transits and short periods, they are well suited for discovery by ground-based surveys such as the Hungarian-made Automated Telescope Network \citep[HATnet;][]{Bakos2004}, HATSouth \citep{Bakos2013}, the Qatar Exoplanet Survey \citep[QES;][]{QES2013}, the Kilodegree Extremely Little Telescope \citep[KELT;][]{KELT2007}, the Wide-Angle Search for Planets \citep[WASP;][]{Pollacco2006}, and the Next-Generation Transit Survey \citep[NGTS;][]{NGTS2018}. As these small-aperture ground surveys are able to detect planets around bright stars, the resulting planet population is useful for follow-up studies of atmospheric composition and structure using occultation and transmission spectroscopy  \citep[e.g.][]{Sing2015, Line2016}.

In this study, we present the discovery of two new hot Jupiter planets: WASP-186b and WASP-187b (TOI-1494.01 and TOI-1493.01). Using Radial Velocity (RV) data from SOPHIE as well as photometry from WASP and TESS, we determine joint probability distributions for the system parameters. In section \ref{sec:observations} we describe the observations of both stars with section \ref{sec:analysis} describing the method used to fit the planet transit and RV data. Finally, section \ref{sec:discussion} provides discussion of the new planets in the context of the known population of exoplanets. 

\section{Observations}\label{sec:observations}
Both WASP-186 and WASP-187 were originally flagged as candidates in the WASP survey data after an initial search of the data using the Box-Least-Squares (BLS) method \citep{Kovacs2002} using the implementation from \cite{CollierCameron2006}.
The planetary nature of both candidates was established with follow-up RV data from the SOPHIE spectrograph \citep{Perruchot2008, Bouchy2009b}. Additionally, both of these planets were recently observed in photometry taken by the Transiting Exoplanet Survey Satellite \citep[TESS;][]{Ricker2015}. A list of all of the observations can be found in Table \ref{tab:observation_list}, with further information found in the following sections. 

\begin{table}
	\centering
	\caption{Summary of available photometric and radial velocity observations for WASP-186 and WASP-187.}
	\label{tab:observation_list}
	\begin{tabular}{lcc}
		\hline
		\hline
		Date & Source & No.obs \\
		\hline
		\multicolumn{3}{l}{\ \ \ \emph{WASP-186b}} \\
		2006 Aug - 2014 Jan & SuperWASP & 81\ 014 \\
		2019 Oct - 2019 Nov & TESS & 821 \\
		2016 Nov - 2017 Nov & SOPHIE HE & 25 \\
		\hline
		\multicolumn{3}{l}{\ \ \ \emph{WASP-187b}} \\
		2004 Jun - 2011 Nov & SuperWASP & 15\ 278 \\
		2019 Oct - 2019 Nov & TESS & 821 \\
		2014 Dec - 2016 Aug & SOPHIE HR & 19 \\
		2016 Sep - 2017 Feb & SOPHIE HE & 13 \\
		\hline
	\end{tabular}
\end{table}

\subsection{WASP Photometry}
\label{sec:photometry} 
WASP-186 was observed by SuperWASP beginning in 2006 and ending in 2014 with a total of 81,014 30-second exposures. The star was originally flagged for further review in July 2016 after inspection of the lightcurve. WASP-187 was observed from 2004 until 2011 with a total of 15,278 observations. In 2014, the target was flagged for further observations. 

The ORCA-TAMTFA transit-search run, which includes multi-season data on 480 fields in the region around the celestial equator that was observed by both SuperWASP and WASP-South, 716 fields from SuperWASP only and 553 from WASP-South only, is the largest and most homogeneous dataset available for WASP. As such, the photometry from this run was used as the basis for analysis by the machine learning method described in \cite{Schanche2019a}. This work combined the results of a Random Forest Classifier and Convolutional Neural Network to find and rank transit candidates. The field in which WASP-186 was identified as a candidate is not part of this run; however, WASP-187 was included and was identified as a good planetary candidate. 


\subsection{SOPHIE spectroscopy}
\label{sec:sophie}

Both WASP-186 and WASP-187 were observed with the SOPHIE spectrograph, first
to establish the planetary nature of the SuperWASP 
transiting candidates, then to determine their masses and orbital eccentricities. 
SOPHIE is dedicated to high-precision RV measurements at the 1.93-m telescope 
of the Haute-Provence Observatory 
\citep{Perruchot2008, Bouchy2009b})
and is widely used for SuperWASP follow-up (e.g. \cite{CollierCameron2007a,Hebrard2013,Schanche2019a}).

WASP-187 was observed between Dec. 2014 and Aug. 2016 in High-Resolution (HR) mode 
with a resolving power $R=76\,000$ and fast readout mode. Further data were taken between Sept. 
2016 and Feb. 2017 in High-Efficiency (HE) mode with a resolving power $R=40\,000$ and 
slow readout mode. HR and HE observations could present a systematic radial velocity shift, so 
they are considered below as independent datasets. WASP-186 was observed  between Nov. 2016 and Nov. 2017, only in HE 
mode and slow readout mode. Two low signal-to-noise observations of WASP-186 were not used. 

The spectra were extracted using the SOPHIE pipeline \citep{Bouchy2009b}
and the radial velocities were measured from the weighted cross-correlation with a 
numerical mask \citep{baranne1996,Pepe2002}. 
They were corrected for the CCD charge transfer inefficiency \citep{Bouchy2009a}
and their error bars were computed from the cross-correlation 
function (CCF) using the method presented by \cite{boisse2010}. 
Following the method described e.g.
in \cite{Pollacco2008} and \cite{Hebrard2008}, 
we estimated and corrected for moonlight contamination by using the second SOPHIE fiber aperture which is targeted on 
the sky while the first aperture points toward the star. The monitoring of constant 
stars revealed no significant instrumental drifts at the epochs of the observations.


The radial velocities are reported in Table~\ref{table_rv}. 
They have larger uncertainties than is typical for SOPHIE due to the rotational line broadening. The resulting CCFs have full width at half maximum of 22 and 21\,km/s 
for WASP-186 and WASP-187, respectively. Still, they show 
significant variations in phase with the SuperWASP transit ephemeris, with semi-amplitudes corresponding 
to companions in the planetary-mass regime.

\begin{figure}
\centering
\includegraphics[width=4.cm, angle=0]{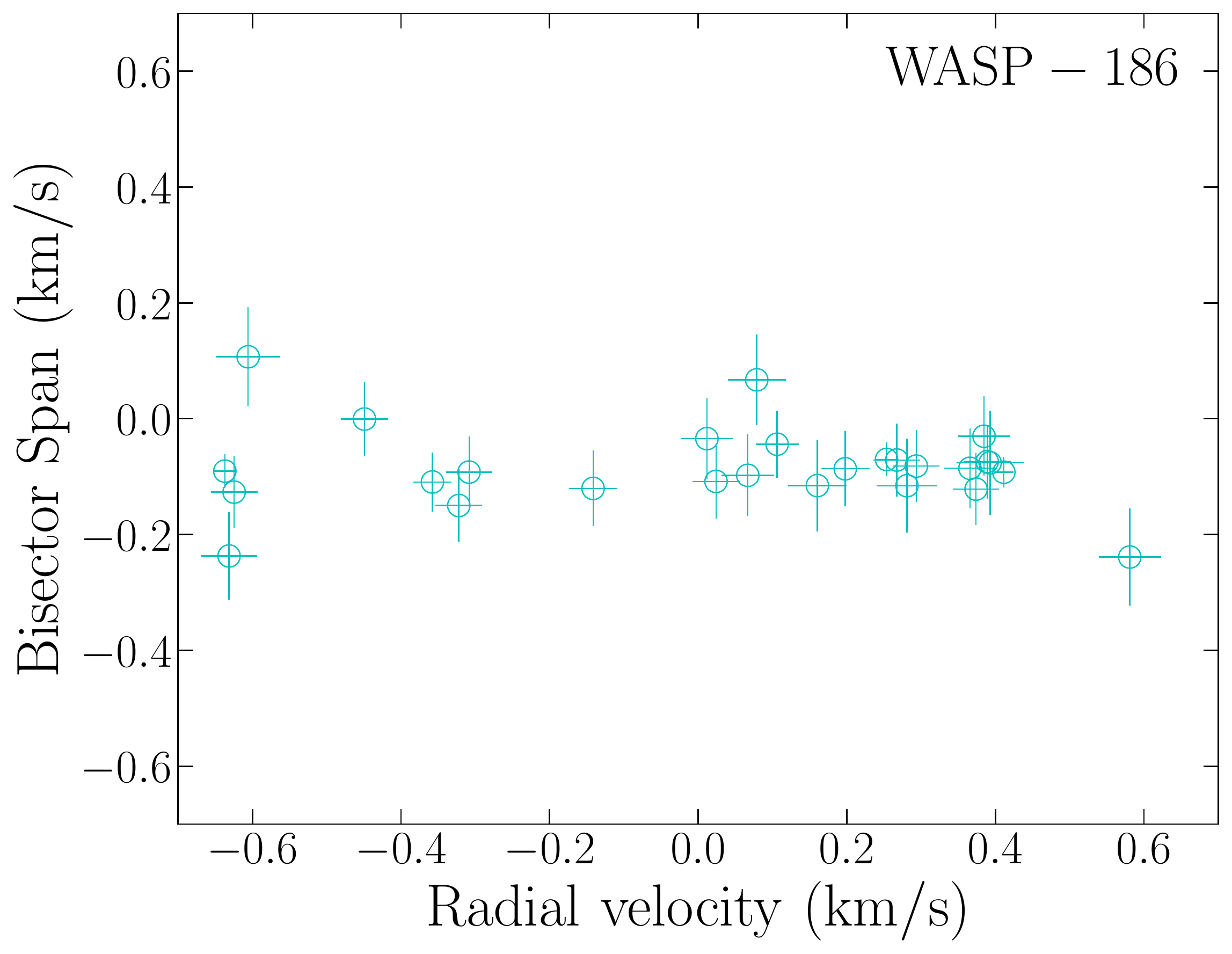}
\includegraphics[width=4.cm, angle=0]{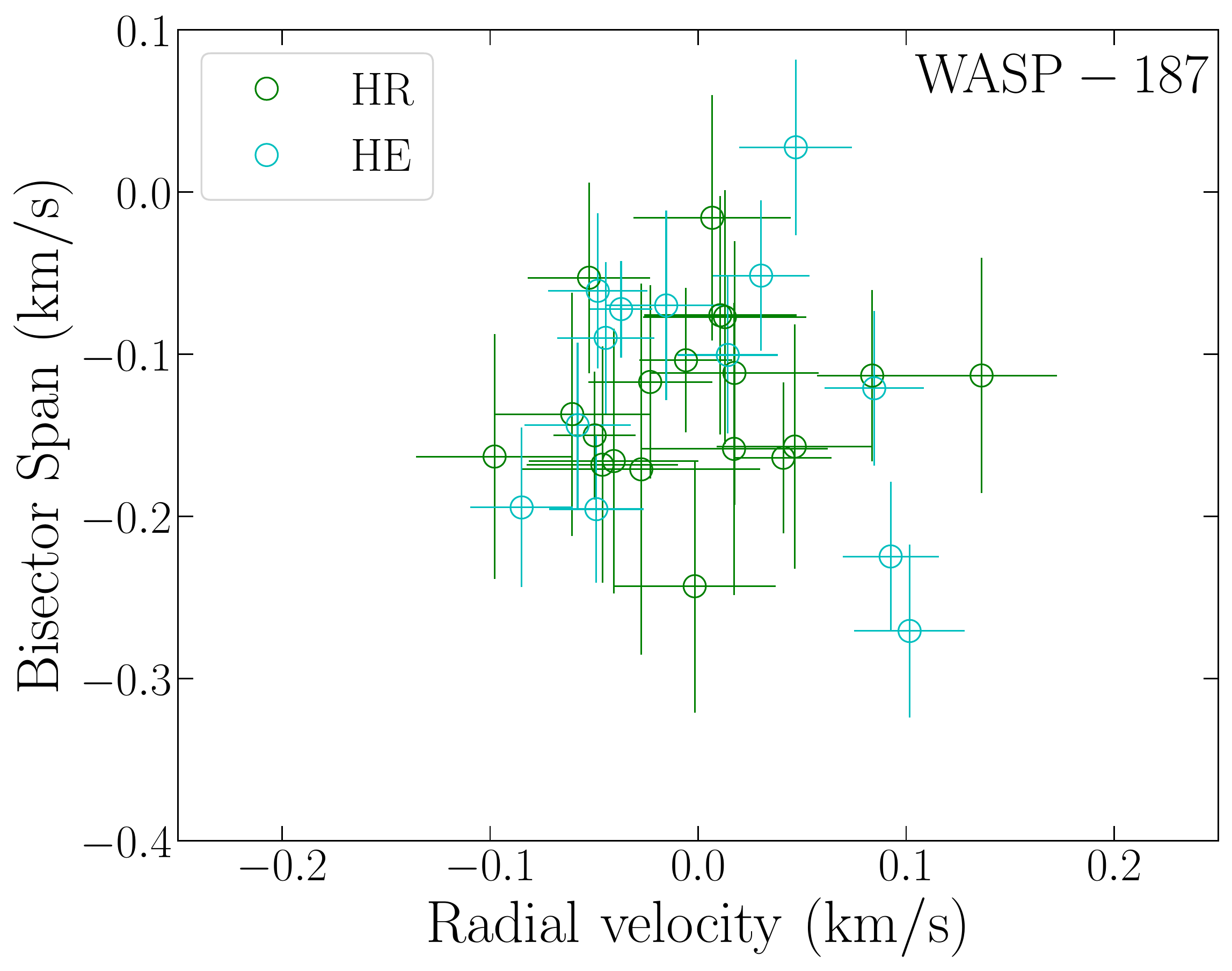}
\caption{Bisector spans as a function of radial velocities for WASP-186 (left) and WASP-187 (right). The radial velocities shown in x-axis are obtained after subtracting the average radial velocity. 
For WASP-187, green and blue circles correspond to the data taken in HE and HR modes, respectively.
Note that the 55.4-m/s fitted shift between HR and HE RVs of WASP-187 is corrected~here. }
\label{fig_bis}
\end{figure}

Radial velocities measured using different stellar masks (F0, G2, K0, or K5) produce 
variations with similar amplitudes, so it is unlikely that these variations are produced by 
blend scenarios composed of stars of different spectral types.
We also checked for signals in the line asymmetry due to blending or magnetic activity using the line bisector measured using the approach of 
\cite{boisse2010}. They are plotted in Fig.~\ref{fig_bis} and show no variations 
correlated with RVs (Pearson correlation coefficients of -0.29 and -0.03 for WASP-186 and 187, respectively).
We can thus conclude the the RV variations are not due to spectral-line profile changes attributable to blends 
or stellar activity, but rather to Doppler shifts due to planetary-mass companions.

\begin{table}
\caption{SOPHIE measurements of the planet-host stars}
\begin{center}
\begin{tabular}{cccrrr}
\hline
\hline
BJD$_{\rm UTC}$ & RV & $\pm$$1\,\sigma$ & bisect.$^\ast$ & exp. & SNR$^\dagger$ \\
-2\,450\,000 & (km/s) & (km/s) & (km/s)  & (sec) &  \\
\hline
\multicolumn{6}{l}{\ \ \ WASP-186 = TOI-1494 (HE mode)} \\
7719.3870 & -5.655 & 0.031 & -0.071 & 257  & 25.3  \\
7721.5162 & -6.064 & 0.033 & -0.120 & 420  & 26.3  \\
7744.3659 & -5.628 & 0.031 & -0.082 & 149  & 25.9  \\
7745.3613 & -6.244 & 0.031 & -0.150 & 165  & 25.7  \\
7746.3741 & -6.230 & 0.031 & -0.092 & 208  & 26.2  \\
7778.2913 & -5.529 & 0.045 & -0.076 & 900  & 19.6  \\
7791.3104 & -6.528 & 0.043 &  0.107 & 1200 & 23.2  \\
7792.2705 & -5.843 & 0.039 &  0.067 & 272  & 25.3  \\
7977.6056 & -6.280 & 0.025 & -0.109 & 900  & 36.0  \\
7988.6111 & -5.669 & 0.015 & -0.070 & 900  & 55.9  \\
7989.6207 & -5.511 & 0.013 & -0.092 & 900  & 61.1  \\
7992.5287 & -6.371 & 0.032 & -0.001 & 229  & 26.0  \\
8003.5848 & -5.762 & 0.040 & -0.115 & 412  & 25.1  \\
8008.6203 & -5.724 & 0.033 & -0.086 & 282  & 25.3  \\
8036.4419 & -5.855 & 0.035 & -0.098 & 382  & 26.4  \\
8037.5082 & -6.559 & 0.014 & -0.090 & 900  & 59.2  \\
8038.4968 & -5.910 & 0.035 & -0.034 & 321  & 26.6  \\
8039.5341 & -5.548 & 0.031 & -0.122 & 196  & 26.1  \\
8040.4692 & -5.533 & 0.032 & -0.074 & 263  & 26.2  \\
8041.4421 & -5.816 & 0.029 & -0.044 & 188  & 29.0  \\
8052.4838 & -6.547 & 0.031 & -0.127 & 147  & 25.9  \\
8053.4492 & -5.898 & 0.032 & -0.108 & 342  & 25.6  \\
8054.4030 & -5.556 & 0.035 & -0.086 & 514  & 25.2  \\
8057.4893 & -6.554 & 0.038 & -0.237 & 441  & 25.5  \\
8085.4759 & -5.537 & 0.035 & -0.030 & 245  & 26.7  \\
\hline
\multicolumn{6}{l}{\ \ \ WASP-187 = TOI-1493 (HR mode)} \\
7020.3536 & -20.287 & 0.029 & -0.144 & 1400 & 37.5  \\
7047.2607 & -20.262 & 0.057 & -0.072 & 1400 & 16.9  \\
7247.6246 & -20.236 & 0.039 & -0.100 & 508  & 26.0  \\
7275.6507 & -20.188 & 0.038 & -0.121 & 407  & 25.7  \\
7303.4805 & -20.284 & 0.020 &  0.028 & 1400 & 45.1  \\
7306.5457 & -20.222 & 0.039 & -0.271 & 708  & 25.1  \\
7309.5569 & -20.217 & 0.041 & -0.061 & 994  & 25.0  \\
7331.5082 & -20.194 & 0.023 & -0.194 & 684  & 39.7  \\
7332.5412 & -20.151 & 0.026 & -0.225 & 628  & 37.2  \\
7333.5243 & -20.258 & 0.030 & -0.051 & 1400 & 33.9  \\
7334.4887 & -20.241 & 0.022 & -0.195 & 655  & 39.4  \\
7335.3682 & -20.281 & 0.036 & -0.090 & 300  & 25.4  \\
7401.2935 & -20.295 & 0.038 & -0.070 & 676  & 25.0  \\
7623.5982 & -20.224 & 0.037 & -0.053 & 384  & 25.1  \\
7624.6495 & -20.217 & 0.045 & -0.171 & 445  & 22.1  \\
7625.6060 & -20.098 & 0.036 & -0.243 & 499  & 24.9  \\
7627.6290 & -20.228 & 0.038 & -0.157 & 446  & 25.3  \\
7628.6165 & -20.333 & 0.038 & -0.150 & 520  & 25.1  \\
7629.5948 & -20.275 & 0.041 & -0.077 & 1619 & 23.1  \\
\hline
\multicolumn{6}{l}{\ \ \ WASP-187 = TOI-1493 (HE mode)} \\
7658.6570 & -20.348 & 0.026 & -0.111 & 638  & 32.6  \\
7659.5249 & -20.327 & 0.015 & -0.164 & 916  & 50.8  \\
7660.4893 & -20.276 & 0.024 & -0.113 & 319  & 32.0  \\
7661.5282 & -20.205 & 0.024 & -0.117 & 281  & 32.1  \\
7681.5194 & -20.243 & 0.027 & -0.104 & 445  & 31.8  \\
7682.5562 & -20.188 & 0.027 & -0.168 & 258  & 32.6  \\
7720.4122 & -20.338 & 0.024 & -0.137 & 277  & 32.3  \\
7721.5065 & -20.375 & 0.025 & -0.076 & 545  & 33.0  \\
7744.3363 & -20.198 & 0.023 & -0.158 & 138  & 32.2  \\
7745.3565 & -20.260 & 0.023 & -0.113 & 171  & 32.0  \\
7746.3691 & -20.339 & 0.023 & -0.016 & 181  & 33.1  \\
7778.2806 & -20.335 & 0.023 & -0.163 & 485 & 33.0  \\
7792.2768 & -20.305 & 0.029 & -0.166 & 263  & 30.3  \\
\hline
\multicolumn{6}{l}{$\ast$: bisector spans; error bars are twice those of the RVs.} \\ 
\multicolumn{6}{l}{$^\dagger$: signal-to-noise ratio per pixel at 550\,nm.} \\
\end{tabular}
\end{center}
\label{table_rv}
\end{table}

\subsection{TESS photometry}
\label{sec:tess}
In October and November 2019, both candidates were observed in Sector 17 of the TESS mission and given the designations as TESS Objects of Interest (TOI)-1494.01 and TOI-1493.01 with four transits observed for each target. However, a momentum dump at the beginning of the observation run coincided with the first transit of WASP-186 leading the affected transit to appear deeper. We therefore remove the first transit from further analysis. Although the event did not disrupt a transit for WASP-187, we remove the data for this time frame to remove any impact on the out-of-transit measurement. Additionally, the data surrounding spacecraft perigee was removed for both lightcurves.x
The Full Frame Images (FFIs) have a cadence of 30 minutes, with each image comprising a stack of 2 second exposures over that time frame. The data were downloaded and the lightcurves were extracted and long term trends were removed using the functions provided in the \textsc{lightkurve} package \citep{Lightkurve2018}. We checked both stars for centroid shifts, but no shifts corresponding to the transits were detected, supporting the conclusion that the transits occur on the target star. 

\section{Analysis} \label{sec:analysis}
\subsection{Spectral Characterization}\label{sec:spectral}
For each star, the SOPHIE spectra not polluted by Moonlight were shifted to a common radial velocity and co-added. The spectral analyses were performed using the process outlined in \cite{Doyle2013}, i.e. the stellar effective temperature ($T_{\rm eff}$) was found using the $H\alpha$ line; the stellar surface gravity (log $g$) was determined from the Na D and Mg b lines; the metallicity was estimated from the width of several Fe I lines; and the projected rotational velocity (v sin i) was determined by convolving the SOPHIE spectrum with the instrumental profile and then using spectrum synthesis to fit the Fe I lines, in agreement with results obtained from the the CCF using the calibration of \cite{boisse2010}. The results of this spectral analysis can be found in Table \ref{tab:spectral}. 

\begin{table}
	\centering
	\caption{Initial stellar parameters from the spectroscopic ($T_{\rm eff}$, log $g$, Fe/H, and $v \sin{i}$), isochrone placement, ($M_{\ast}$ and Age), and {\em Gaia}+IRFM ($\varpi$, $R_{\ast}$) analysis of WASP-186 and WASP-187. }
	\label{tab:spectral}
	\begin{tabular}{lll}
		\hline
		\hline
		Parameter & WASP-186 & WASP-187 \\
		\hline
		$T_{\rm eff}$ (K) & $6300 \pm 100$ & $6100 \pm 100$ \\
		\noalign{\smallskip}
		log $g$ & $4.1 \pm 0.2$ & $3.8 \pm 0.2$ \\
		\noalign{\smallskip}
		Fe/H & $-0.08 \pm 0.14$ & $0.0 \pm 0.11$ \\
		\noalign{\smallskip}
		$v \sin{i}$ (${\rm km\,s}^{-1}$) & $15.6 \pm 0.9$ & $15.3 \pm 1.0 $ \\
		\noalign{\smallskip}
		$M_{\ast}$ & $1.21^{+0.07}_{-0.08}$ & $1.53^{+0.07}_{-0.09}$ \\
		\noalign{\smallskip}
		Age (Gyr) & $3.1^{+1.0}_{-0.8}$ & $2.55^{+0.49}_{-0.25}$ \\
		\noalign{\smallskip}
		Parallax $\varpi$ (mas) & 3.563 $\pm{0.042}$ & 2.667 $\pm{0.048}$ \\
		\noalign{\smallskip}
        $R_{\ast}$ ($R_{\odot}$) & 1.46 $\pm{0.02}$ & 2.87 $\pm{0.05}$\\
		\hline
	\end{tabular}
\end{table}

\subsection{Gaia IRFM}\label{sec:GaiaIRFM}
The infrared flux method (IRFM), introduced by \cite{Blackwell1977}, is a semi-direct way to measure stellar angular diameter and effective temperature. This method combines flux measurements at different wavelengths with stellar atmospheric models to determine the stellar properties. The IRFM method has been implemented by several groups, e.g. \citep{ Alonso1994, Ramirez2005,Hernandez2009, Casagrande2010}. Here we expand the method with the incorporation of data from \textit{Gaia}. 

The magnitudes and corresponding uncertainties for WASP-186 and WASP-187 in the \textit{Gaia} G, $G_{\rm BP}$ and $G_{\rm RP}$ bandpasses \citep{Riello2018} were retrieved from the second data release (DR2) archive \citep{Gaia2018}, along with data taken in the J, H, and K filters from the 2MASS survey \citep{2mass} and in the W1 and W2 (3.4 and 4.6 $\mu$m) bandpasses from the WISE survey \citep{WISE}. This information is used together with the stellar synthetic spectra atlas of \cite{Castelli2003} to find initial estimates of stellar effective temperature and angular diameter ($\theta$). The \textit{Gaia} parallax measurement ($\varpi$) is then used to produce a radius measurement for the star. 

\subsection{Stellar Masses}\label{sec:masses}
 Using stellar parameters derived from the spectral line fitting (shown in Table~\ref{tab:spectral}), stellar masses and ages for WASP-186 and WASP-187 were determined by the isochrone placement functionality of the {\sc mcmci} tool \citep{Bonfanti2020}. Following an MCMC approach, with five chains of 100\,000 steps and a burn-in fraction of 20\,per cent, stellar masses were calculated by interpolating over grids of stellar isochrones and evolutionary tracks to be $M_{\rm *} = 1.21_{-0.08}^{+0.07}$\,$M_{\odot}$ and $M_{\rm *} = 1.53_{-0.09}^{+0.07}$\,$M_{\odot}$ for WASP-186 and WASP-187, respectively. The corresponding stellar ages were found to be $3.1_{-0.8}^{+1.0}$ and $2.55_{-0.25}^{+0.49}$ Gyr. This places WASP-187 near the main-sequence turn-off on evolutionary tracks for the most probable range of masses. 

\subsection{Planetary Parameters}\label{sec:mcmc}
The MCMC approach is commonly used in exoplanet model fitting as it is efficiently able to fit a model to the data while at the same time providing a posterior probability distribution of each fitted parameter. The implementation used here is modeled on \cite{CollierCameron2007b}. The code aims to fit the stellar parameters along with the transit and RV data simultaneously. 

The jump parameters used to describe the data are chosen to be independent of each other, therefore the fit of the data to the model relies on the transition of the MCMC jump parameters to the physical variables (For a full description of the transition from these jump parameters to physical variables, see Appendix \ref{app:State2Phys}). The jump parameters used here are: the transit epoch ($T_c$), period ($P$), impact parameter ($b$), transit width ($w$), transit depth ($(R_P/R_S)^2$), received flux ($f$), parallax ($\varpi$), stellar radius ($R_{\ast}$), extinction ($E(B{-}V)$), log of the system error which accounts for zero-point uncertainties in the definitions of the flux-to-magnitude conversions for the different bandpasses in the IRFM calculation ($\log{\sigma_{\rm sys}}$), RV amplitude ($K$), RV offset ($\gamma$), and RV jitter ($\sigma_{\rm jit}$). For WASP-187 the RV offset and jitter are treated separately for the High-Resolution and High-Efficiency modes. The eccentricity $e$ and argument of periastron $\omega$ are parameterized as $\sqrt{e}\cos{\omega}$ and $\sqrt{e}\sin{\omega}$; the fit to the orbit of WASP-187 was, however, found to be consistent with an eccentricity of 0. Therefore the final MCMC held these values constant. Finally, the transit depth of WASP-187 was underestimated when WASP data were included in the fit. There are no contaminating stars within 114 arcsec. The most likely explanation is a slight dilution of the signal from the detrending of the WASP data. To account for the dilution, we fit an additional positive, constant flux offset to the WASP fluxes for this star, thereby allowing the TESS data to dominate the depth determination while retaining the timing information provided by WASP.

Initial fits for the depth, width, impact parameter, period, and epoch for the photometric datasets were done using the Transit Model in the \textsc{\tt PyCHEOPS} v0.6.0 Python package\footnote{\url{https://github.com/pmaxted/pycheops}}. The power-2 limb darkening coefficients \citep{Maxted2018} are interpolated from tables for TESS and WASP separately for the initial fit, as well as at every step in the MCMC. The initial fit for the radial velocity parameters were obtained with the \textsc{RadVel} package. 

Several additional pieces of prior information are incorporated into the model. Given $T_{\rm eff}$, $\log g$, and Fe/H, the stellar surface flux spectrum is computed and attenuated by a galactic extinction law characterised by $E(B{-}V)$. The resulting reddened spectrum is folded through the \textit{Gaia}, 2MASS and WISE photon-weighted filter transmission curves and scaled by the zero-points and the squared angular radius to obtain synthetic apparent magnitudes. The residuals of the observed minus synthetic magnitudes in the eight bandpasses and their uncertainties contribute directly to the likelihood at each jump. There is also a prior imposed on the parallax from the measurement in \textit{Gaia} DR2 ($3.562 \pm 0.043 $ and $2.664 \pm 0.044$ mas for WASP-186 and WASP-187 respectively). $T_{\rm eff}$ and $\log g$ priors are imposed from the spectroscopic analysis described in section \ref{sec:spectral}. Finally, the stellar mass prior was determined from the isochrone placement method described in section \ref{sec:masses}.

\begin{figure}
\centering
\includegraphics[width=8.cm, angle=0]{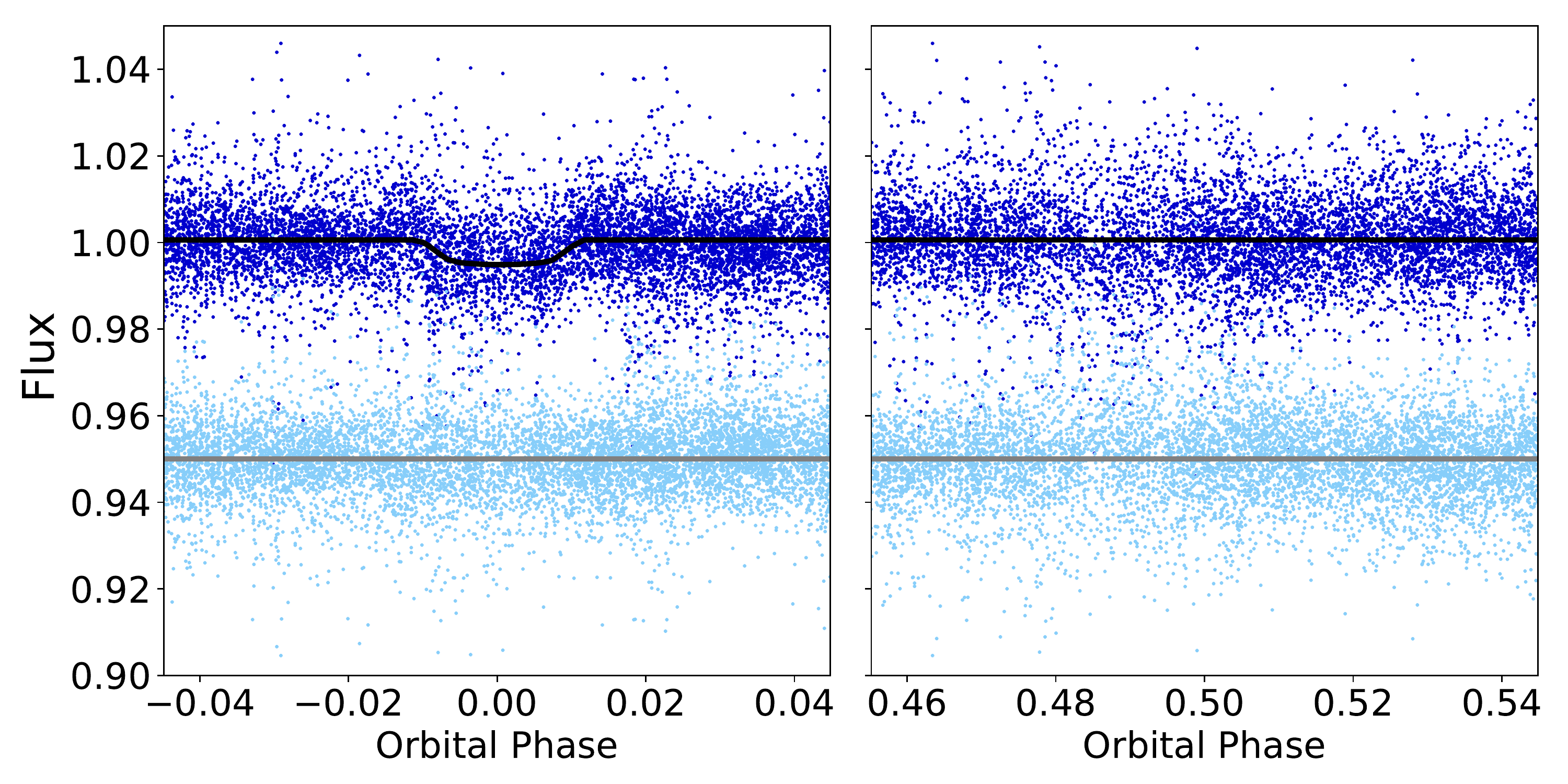}
\includegraphics[width=8.cm, angle=0]{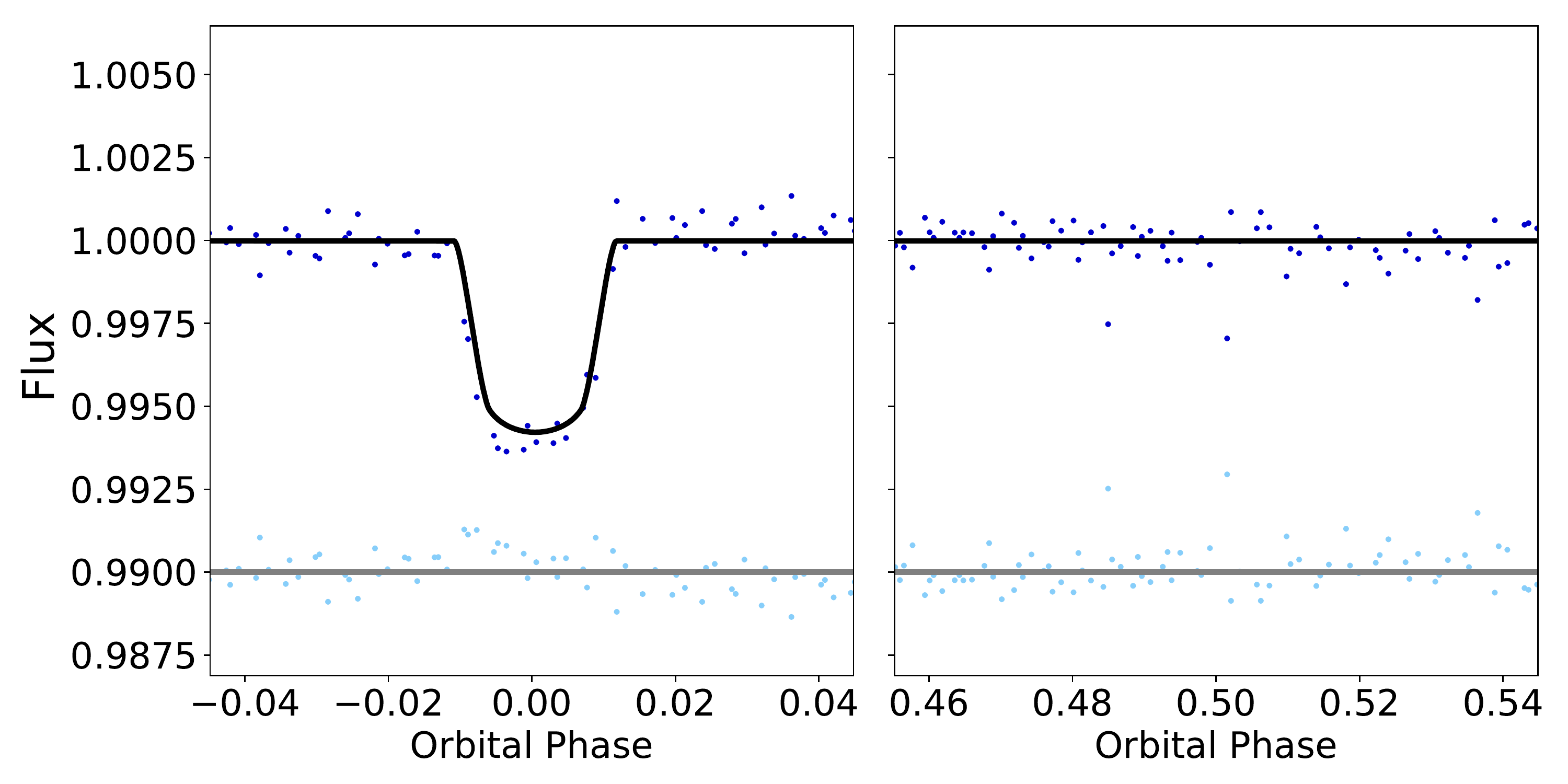}
\includegraphics[width=8.cm, angle=0]{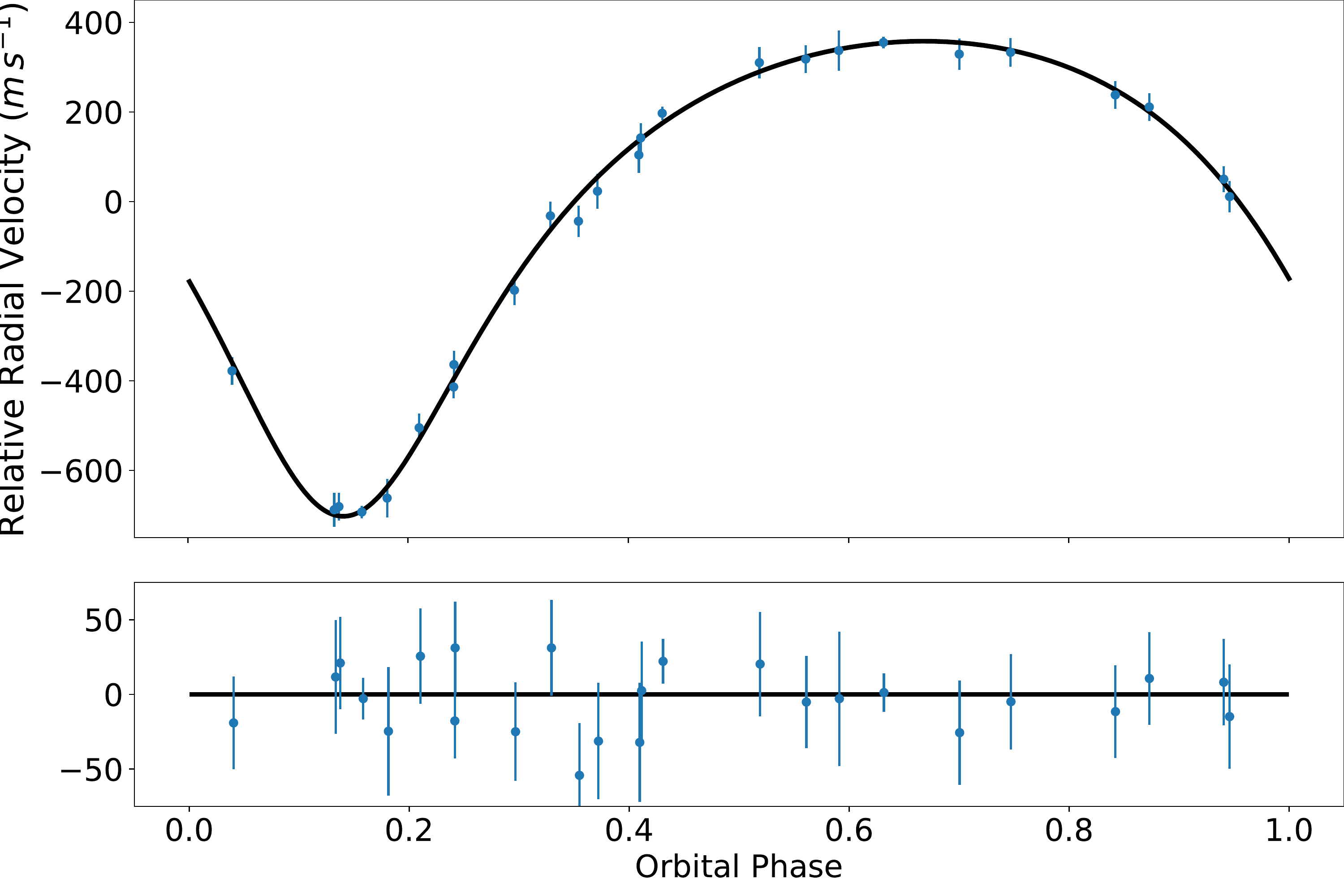}
\caption{WASP (top), TESS (middle), and SOPHIE (bottom) data for WASP-186b phase folded on the best-fitting period. Residuals to the fit are shown below the data. }
\label{fig:WASP186_fit}
\end{figure}

\begin{figure}
\centering
\includegraphics[width=8.cm, angle=0]{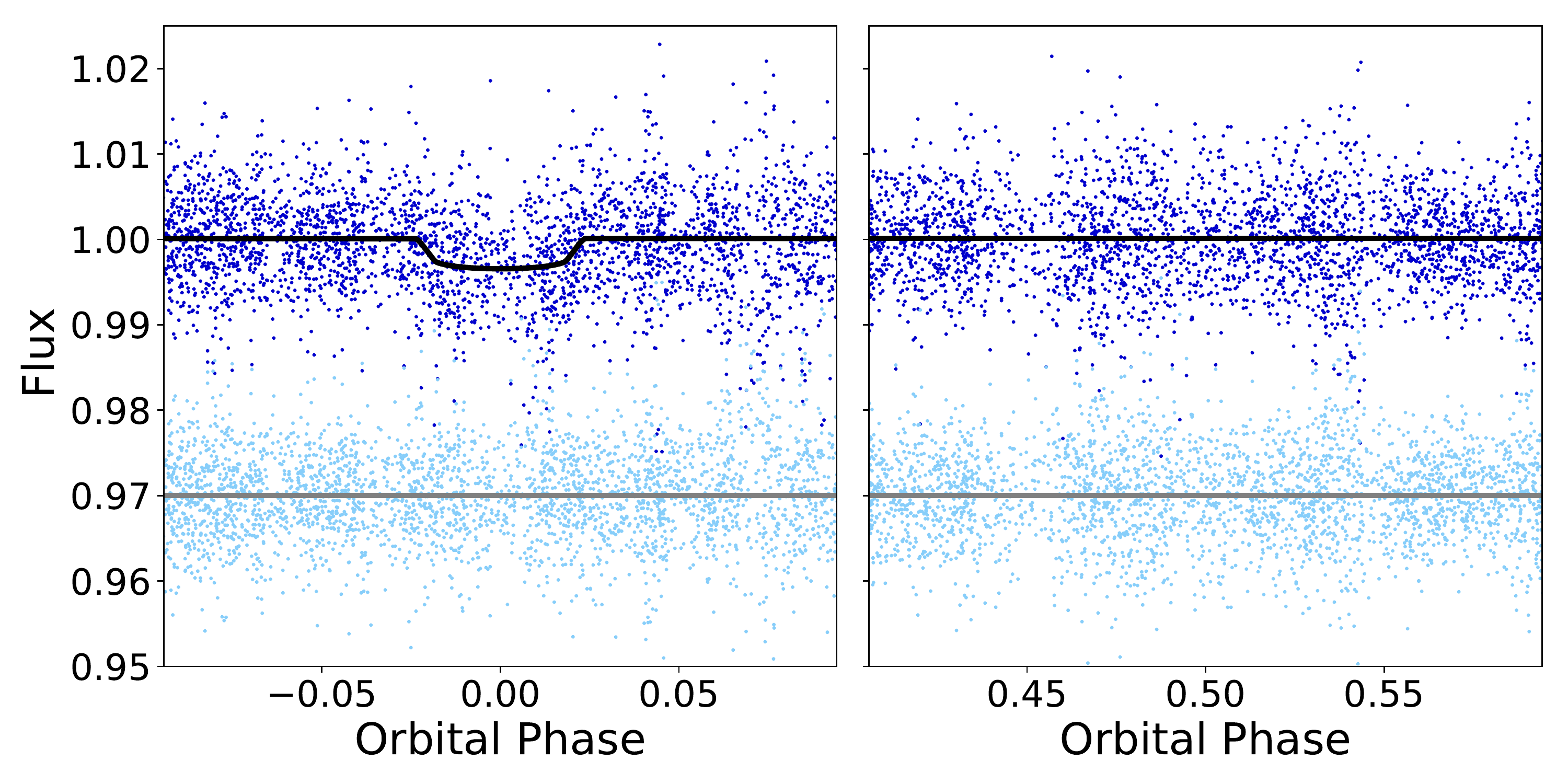}
\includegraphics[width=8.cm, angle=0]{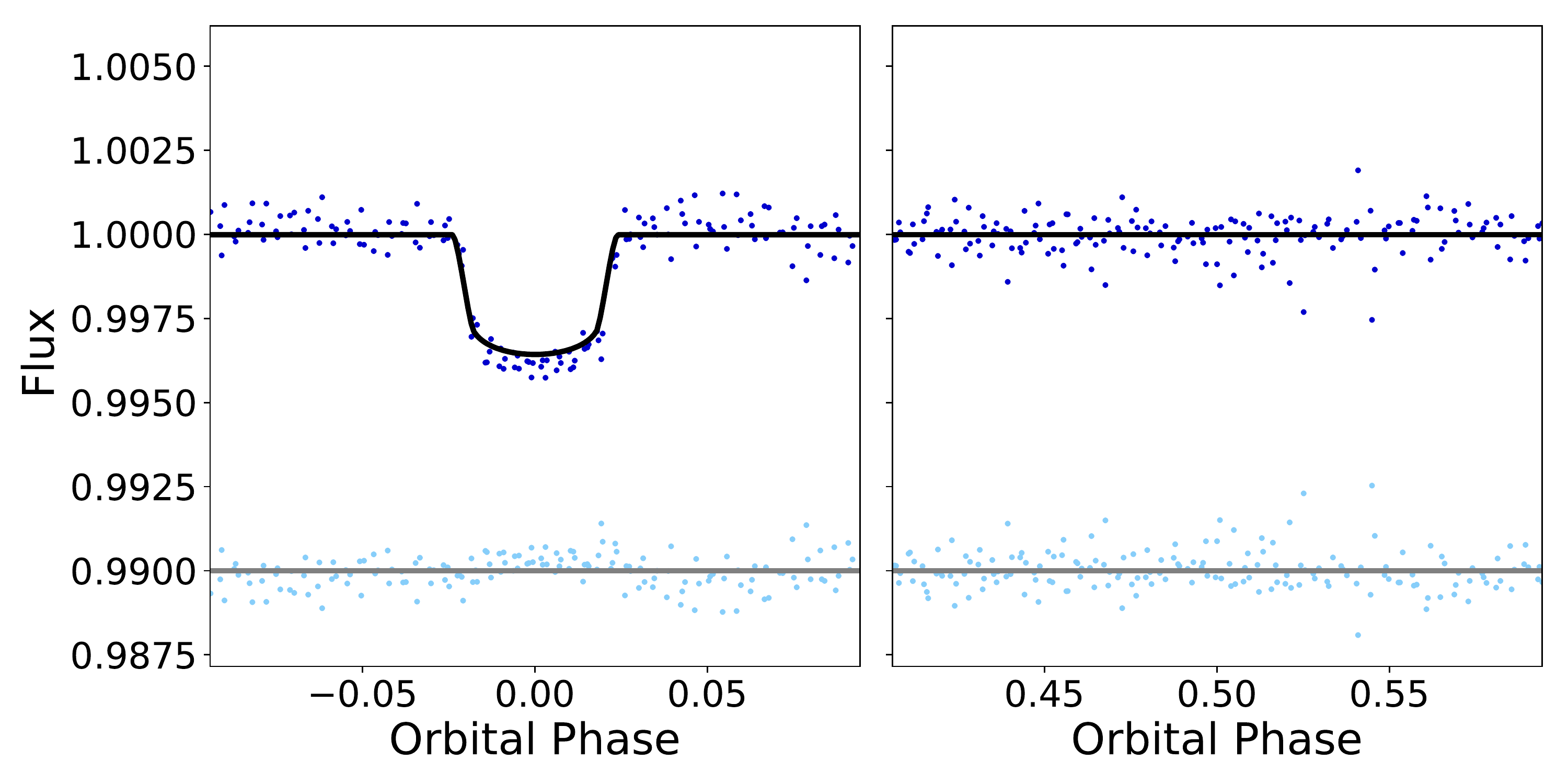}
\includegraphics[width=8.cm, angle=0]{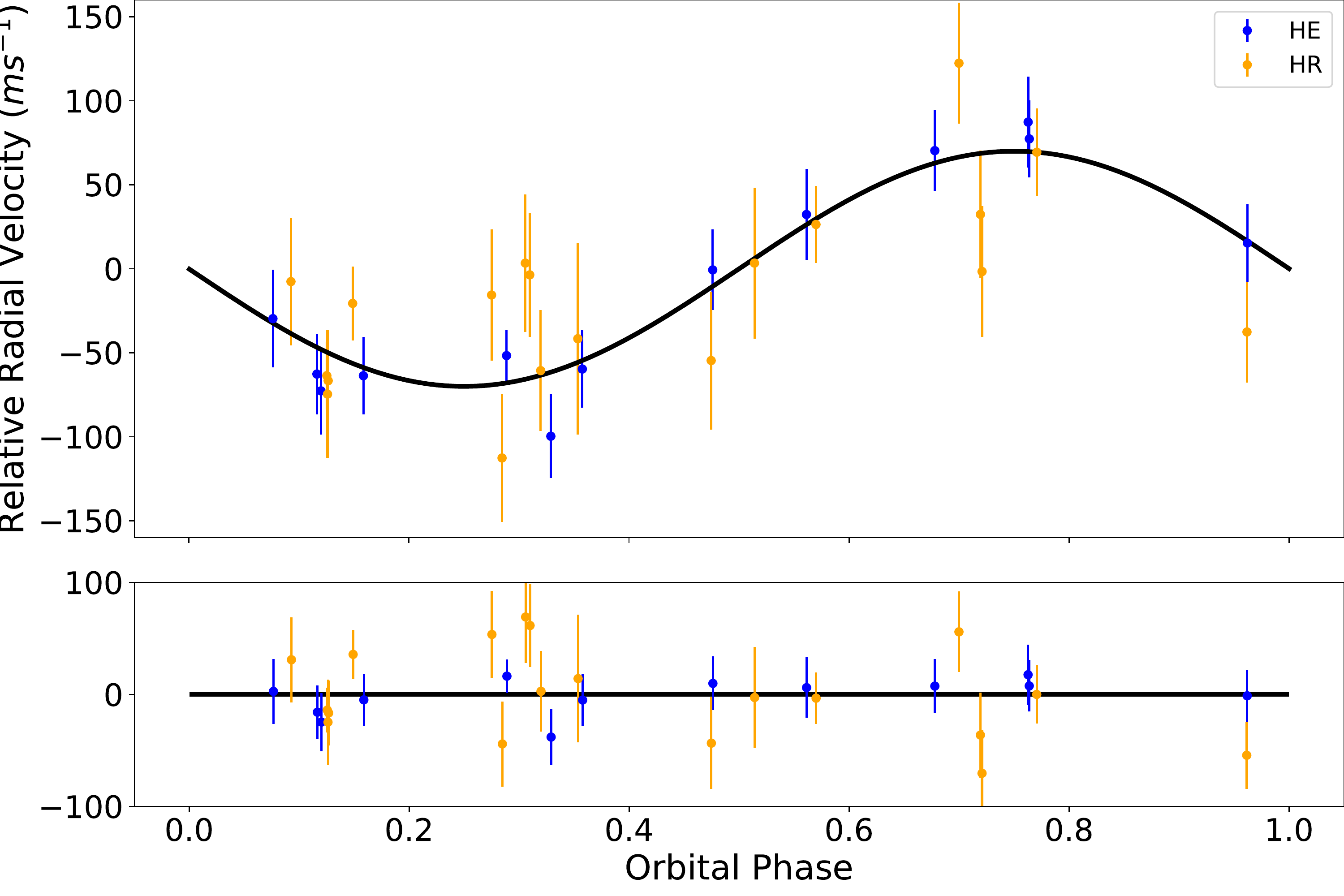}
\caption{WASP (top), TESS (middle), and SOPHIE (bottom) data for WASP-187b phase folded on the best-fitting period. Residuals to the fit are shown below the data. }
\label{fig:WASP187_fit}
\end{figure}

The MCMC went through 3 burn-in phases of 6\ 000, 2\ 000, and 2\ 000 steps, with updates of the jump lengths after each phase. The final MCMC was run for 50\ 000 steps. The final corner plots for the MCMC runs are shown in Figures \ref{fig:corner_plot_WASP186_photo} through \ref{fig:corner_plot_WASP187_stellar} in the appendix. The resulting best-fit model for WASP-186b and WASP-187b are shown in Figures \ref{fig:WASP186_fit} and \ref{fig:WASP187_fit}.

\begin{table*}
\caption{System parameters for WASP-186 and WASP-187}
\begin{center}
\begin{tabular}{lccc}
\hline
\hline
Parameters & Symbol (Unit) & WASP-186 & WASP-187 \\
\hline
\multicolumn{4}{l}{\emph{\textbf{Stellar Parameters}}} \\
WASP ID & & 1SWASP J011558.85+213700.9 & 1SWASP J010953.96+254054.0 \\
\noalign{\smallskip}
TESS ID & & TOI-1494.01/TIC-411608801 & TOI-1493.01/TIC-15692883 \\
\noalign{\smallskip}
Gaia ID & & 2790691147020786816 & 306410392895767680 \\
\noalign{\smallskip}
Right Ascension & RA (hh:mm:ss) & 01:15:58.85 & 01:09:53.96 \\
\noalign{\smallskip}
Declination & Dec (dd:mm:ss) & +21:37:00.9 & +25:40:54.0 \\
\noalign{\smallskip}
Visual Magnitude & Vmag (mag) & 10.82 & 10.30 \\
\noalign{\smallskip}
TESS Magnitude & Tmag (mag) & 10.30 & 9.71 \\
\noalign{\smallskip}
Gaia Magnitude & Gmag (mag) & 10.65 & 10.13 \\
\noalign{\smallskip}
Stellar Mass & $M_{\ast}$ ($M_{\odot}$) & $1.22^{+0.07}_{-0.08}$ & $1.54 \pm 0.09$ \\
\noalign{\smallskip}
Stellar Radius & $R_{\ast}$ ($R_{\odot}$) & $1.47 \pm 0.02$ & $2.83 \pm 0.05$ \\
\noalign{\smallskip}
Effective Temperature & $T_{\rm eff}$ (K) & $6361^{+105}_{-82}$ & $6150^{+92}_{-85}$\\
\noalign{\smallskip}
Parallax & $\varpi$ (mas) & $3.571^{+0.044}_{-0.042}$ & $2.663^{+0.046}_{-0.043}$ \\
\noalign{\smallskip}
Stellar Density & $\rho_s$ ($\rho_{\odot}$) & $0.387^{+0.028}_{-0.027}$ & $0.068 \pm 0.005$ \\
\noalign{\smallskip}
Surface Gravity & log $g$ \ (cgs) & $4.193^{+0.028}_{-0.029}$ & $3.722 \pm 0.029$ \\
\noalign{\smallskip}
Received Flux & f*1e-9 (cgs) &$1.266^{+0.007}_{-0.006}$ & $2.055 \pm 0.008$ \\
\noalign{\smallskip}
Extinction & $E(B{-}V)$ (mag) & $0.03 \pm 0.02$ & $0.09 \pm 0.02$ \\
\hline
\multicolumn{4}{l}{\emph{\textbf{Planet Parameters}}} \\
Period & P (d) & $5.026799^{+.000012}_{-.000014}$ & $5.147878^{+.000005}_{-.000009}$ \\
\noalign{\smallskip}
Transit Epoch & Tc-2450000 & $6237.1195 \pm 0.0009$ & $5197.3529^{+0.002}_{-0.0022}$ \\
\noalign{\smallskip}
Transit Width & $w$ (hr) & $2.704^{+0.048}_{-0.051}$ & $5.82^{+0.095}_{-0.091}$ \\
\noalign{\smallskip}
Transit Depth & $(R_P/R_S)^2$ & $0.0061 \pm 0.0003$ & $0.0035 \pm 0.0002$ \\
\noalign{\smallskip}
Planet Mass & $M_p \ (M_{Jup})$ & $4.22 \pm 0.18$ & $0.8 \pm 0.09$ \\
\noalign{\smallskip}
Planet Radius & $R_p \ (R_{Jup})$ & $1.11 \pm 0.03$ & $1.64 \pm 0.05$ \\
\noalign{\smallskip}
Semi-major Axis & a (au) & $0.06^{+0.0012}_{-0.0013}$ & $0.0653 \pm 0.0013$ \\
\noalign{\smallskip}
Impact Parameter & $b$ & $0.84^{+0.01}_{-0.02}$ & $0.76 \pm 0.02$\\
\noalign{\smallskip}
Orbital Eccentricity & $e$ & $0.33 \pm 0.01$ & 0 (Fixed) \\
\noalign{\smallskip}
Argument of Periastron & $\omega$ & $3.02^{+0.05}_{-0.06}$ & 0 (Fixed) \\
\noalign{\smallskip}
Planet Density & $\rho_p \ (\rho_{Jup})$ & $2.881^{+0.3}_{-0.279}$ & $0.169^{+0.026}_{-0.023}$ \\
\noalign{\smallskip}
Surface Gravity & log $g$ (cgs) & $3.93 \pm 0.03$ & $2.87^{+0.05}_{-0.06}$ \\
\noalign{\smallskip}
RV Semi-amplitude & K ($m\ s^{-1}$) & $530.29^{+8.43}_{-8.59}$ & $70.0^{+6.84}_{-7.75}$\\
\noalign{\smallskip}
Zero-point Uncertainty & $\log{\sigma_{\rm sys}}$ & $-2.19^{+0.41}_{-0.34}$ & $-2.16^{+0.32}_{-0.25}$ \\
\noalign{\smallskip}
RV offset & $\gamma$ ($m\ s^{-1}$) & $-5866.2^{+5.7}_{-5.6}$ & $-20275.3^{+6.9}_{-6.8}$ (HE), $-20220.4^{+9.2}_{-9.3}$ (HR) \\
\noalign{\smallskip}
RV Jitter & $\sigma_{\rm jit}$ ($m\ s^{-1}$ & $0.8^{+7.1}_{-7.3}$ & $3.6^{+4.8}_{-2.8}$ (HE), $17.0^{+12.7}_{-8.7}$ (HR) \\
\hline
\end{tabular}
\end{center}
\label{tab:results_table}
\end{table*}


\section{Discussion and Conclusions}\label{sec:discussion}
The full set of parameters derived for WASP-186b and WASP-187b can be found in Table \ref{tab:results_table}. Note that the values reported for $T_{\rm eff}$, stellar log $g$, $\varpi$, 
$M_{\ast}$ and $R_{\ast}$ were determined by the MCMC analysis and differ slightly from the prior values shown in Table \ref{tab:spectral}.

\begin{figure}
\centering
\includegraphics[width=8cm, angle=0]{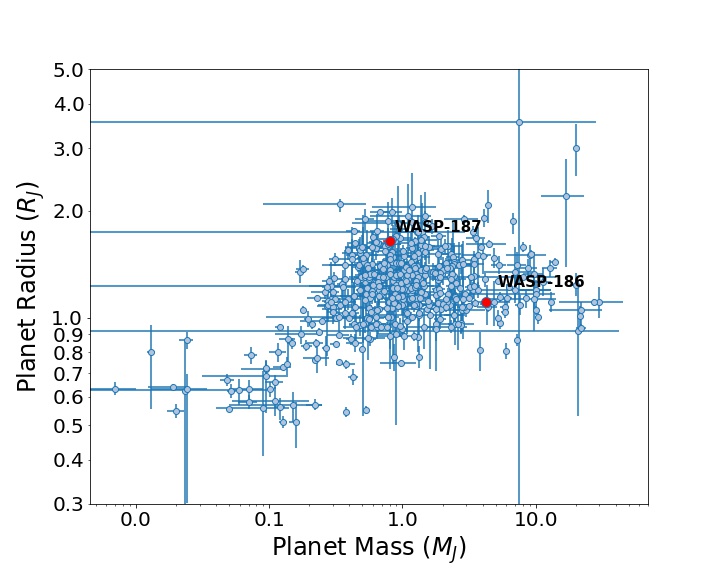}
\caption{Planet mass versus planet radius for all Jupiter-sized planets ($R_P > 0.5R_J$) with mass and radius measurements. Data for this and subsequent plots was obtained from the NASA Exoplanet Archive  http://exoplanetarchive.ipac.caltech.edu. }
\label{Fig:Mp_Rp_plot}
\end{figure}

WASP-186 is a mid-F type star with an effective temperature of $6361^{+105}_{-82}\,$K, agreeing with the spectral determination within errors. The parallax estimate corresponds to a distance of 280.71$^{+13.05}_{-11.96}$ pc \citep{BailerJones2018}. The star is rotating with a $v \sin{i}$ of $15.6 \pm{0.9}\, {\rm km\,s}^{-1}$. As can be seen in Figure \ref{Fig:Mp_Rp_plot}, the planet has a radius typical for a hot Jupiter ($1.11 \pm{0.03}\, R_ {\rm Jup}$), but is quite massive at $4.22 \pm{0.18}\, M_ {\rm Jup}$. WASP-186b therefore fits among the most massive and dense hot Jupiters known. WASP-186b is also notable as the orbit has an eccentricity of 0.33 $\pm{0.01}$, pointing to late-time high-eccentricity migration, rather than disc migration \citep{Rasio1996, Ford2008}. Using equation 1 of \cite{Dobbs-Dixon2004} and an estimate for $Q'_p$ of 10$^6$ in line with the estimation from \cite{Yoder1981}, the time scale of eccentricity damping via tidal disturbance is on the order of 15.7 Gyr, well above the estimated age of 3.1$^{+1.0}_{-0.8}$ Gyr. WASP-186b therefore joins the small group of massive and eccentric planets, including WASP-150b \cite{Cooke2020}, WASP-162b \citep{Hellier2019}, HATS-41b \citep{Bento2018}, XO-3b \citep{Johns-Krull2008}, and HAT-P-2b \citep{Bakos2007}. Finally, the planet has an equilibrium temperature ($T_{eq}$) of $1348^{+23}_{-22}$ K, assuming zero albedo and isotropic blackbody re-radiation. 

\begin{figure}
\centering
\includegraphics[width=8cm, angle=0]{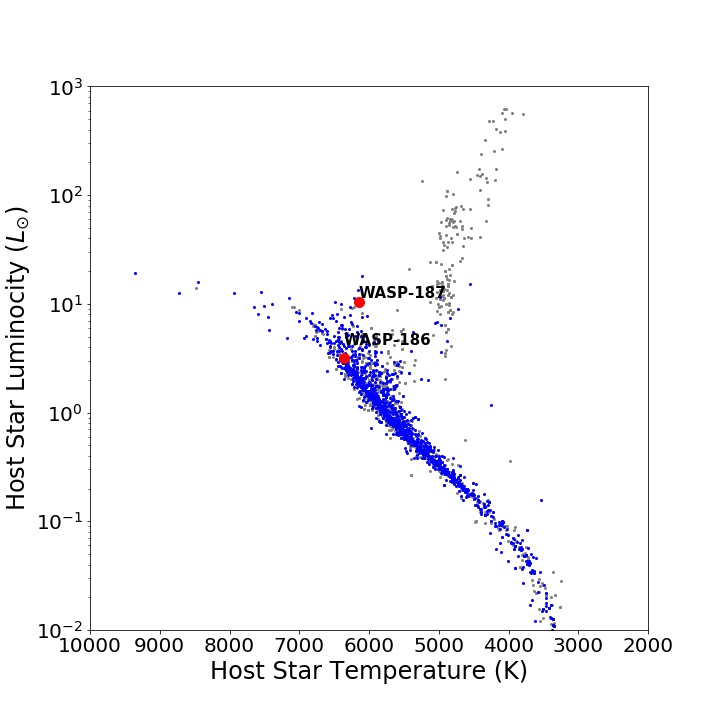}
\caption{H-R diagram showing temperature versus stellar luminosity for all stars known to host exoplanets. Stars hosting planets with periods less than 10 days are shown in blue, while stars hosting planets with longer periods are in gray. Note that WASP-187 lies to the right of the main sequence, indicating that the star has entered the post-main sequence phase of its life.  }
\label{fig:HRdiagram}
\end{figure}

\begin{figure}
\centering
\includegraphics[width=8cm, angle=0]{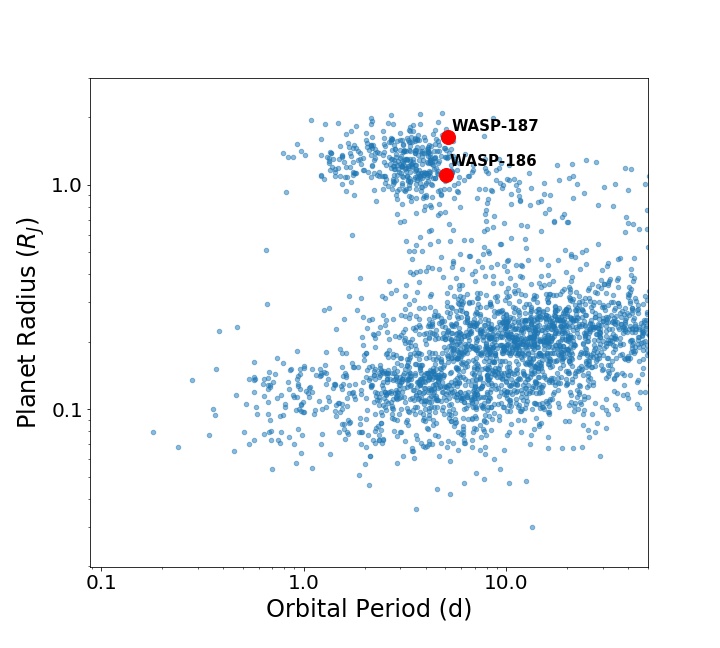}
\caption{Orbital period versus planet radius for all known exoplanets with a period less than 50 days. }
\label{fig:Per_Rp}
\end{figure}

The host star for WASP-187b has begun to evolve away from the main sequence (see Fig. \ref{fig:HRdiagram}), indicated by the stellar effective temperature of $6150^{+92}_{-85}\,$K, mass of $1.54 \pm{0.09}\, M_{\odot}$, and radius of $2.83 \pm{0.05}\, R_{\odot}$. The star is 375.52$\pm{.45}$ pc away \citep{BailerJones2018} and has a projected rotation is similar to that of WASP-186 at 15.3 $\pm{1.0}\,  {\rm km\,s}^{-1}$, indicating the rotation has slowed since leaving the main sequence \citep{Wolff1997}. WASP-187b is hotter ($T_{eq} = 1726^{+31}_{-29}$) and significantly less dense than WASP-186b with a mass of $0.80\pm{0.09}\, M_ {\rm Jup}$ and a radius of $1.64\pm{0.05}\, R_ {\rm Jup}$, suggesting that this planet could be undergoing re-inflation \citep{Hartman2016}. Both planets have a period of around 5 days, but their radii lie near the low and high radius boundaries for known hot Jupiters of that period (see Fig. \ref{fig:Per_Rp}). 

\section*{Acknowledgements}
NS acknowledges the support of NPRP grant \#X-019-1-006 from the Qatar National Research Fund (a member of Qatar Foundation). ACC acknowledges support from the Science and Technology Facilities Council (STFC) consolidated grant number ST/R000824/1 and UKSA grant ST/R003203/1. CAH and UCK are supported by STFC under Consolidated Grant ST/T000295/1. DLP, RGW and PJW have been supported by STFC consolidated grants ST/P000495/1 and ST/T000406/1. This work was supported by the Swiss National Science Foundation (SNSF). This paper includes data collected by the TESS mission. Funding for the TESS mission is provided by the NASA Explorer Program. This research made use of Lightkurve, a Python package for Kepler and TESS data analysis (Lightkurve Collaboration, 2018). This research has made use of the NASA Exoplanet Archive, which is operated by the California Institute of Technology, under contract with the National Aeronautics and Space Administration under the Exoplanet Exploration Program.

\section*{Data Availability}
The data used in this publication can be accessed at https://doi.org/10.17630/c311756f-557e-4955-b50e-980633ded8f9




\bibliographystyle{mnras}
\bibliography{papers} 













\newpage

\appendix 
\section{State to Physical Variables} \label{app:State2Phys}
In order to convert the MCMC jump parameters ($P$, $w$, $d$, $b$, $\sqrt{e}\cos{\omega}$, $\sqrt{e}\sin{\omega}$, $f$, $\varpi$, $R_{\ast}$, $E(B{-}V)$, and $\log(\sigma_{\rm sys})$) to their physical meaning for the system, the following equations were used:

The eccentricity $e$ and argument of periastron $\omega$ are found using:

\begin{equation}
    e = (\sqrt{e}\cos{\omega})^2 + (\sqrt{e}\sin{\omega})^2
\end{equation}
and 
\begin{equation}
    \omega = \tan^{-1}(\sqrt{e}\sin{\omega}, \sqrt{e}\cos{\omega})
\end{equation}

The transit duration (in days) that would be expected if the impact parameter were 0 is given by

\begin{equation}
    w_0 = \dfrac{w (1+k)}{\sqrt{(1+k)^2-b^2}}
\end{equation}

where $w$ is a proxy for the transit width and $k$ is the ratio of the planetary to stellar radii, $R_p/R_{\ast}$. The ratio of the stellar radius to the semi-major axis is then given by

\begin{equation}
    \dfrac{R_{\ast}}{a} = \dfrac{\pi \  w_0 \ (1+e\sin{\omega})}{(1+k)  P  \sqrt{(1-e^2)}}
\end{equation}
where P is the orbital period in days. As described by \cite{Winn2009}, $\cos{i}$ and $\sin{i}$ are then calculated as 

\begin{equation}
    \cos{i} = b\dfrac{R_{\ast}}{a}\dfrac{1+e\sin{\omega}}{(1-e^2)} \\
    \sin{i} = \sqrt{1-\cos{i}^2}
\end{equation}

The next step is to get the angular radius of the star. First the radius and parallax of the star are used to get the angular radius ($\theta \equiv \dfrac{R_{\ast}}{d}$):

\begin{equation}
    \theta = \dfrac{R_{\ast} \ (\varpi+0.082)}{1000 \star 180 \star 3600}
\end{equation}
where $\varpi$ is the parallax in milli-arcseconds, and the correction of \cite{Stassun2018} is applied to the parallax. 

The orbital separation can then be found by:

\begin{equation}
    \dfrac{a}{a_{\oplus}} = \dfrac{\theta}{\varpi}\dfrac{a}{R_{\ast}}
\end{equation}

with $a_{\oplus}$ being one astronomical unit and $\varpi$ expressed in radians.

The mass of the star can then be calculated using Kelper's third law and appropriate unit conversions:
\begin{equation}
    \dfrac{M_{\ast}}{M_{\odot}} = \left(\dfrac{a}{a_{\oplus}}\right)^3 \left(\dfrac{P}{ 1{\rm year}}\right)^{-2}
\end{equation}

The angular radius is also utilized to find the stellar surface flux $F$:

\begin{equation}
    F = f / \theta^2.
\end{equation}

The stellar surface flux is corrected for extinction iteratively:

\begin{equation}
    F_{\rm ext} = F \ 10^{0.4 \ E(B{-}V) \ R_{\rm bol}}.
\end{equation}

$R_{\rm bol}$ used here is the bolomentric extinction to reddening ratio, determined by optimizing the fit to stellar radii in the asteroseismic samples provided in \cite{SilvaAguirre2015}:

\begin{equation}
    R_{\rm bol} = 2.31 + 0.000509(T_{\rm eff} - 5894)
\end{equation}

The flux is then used to find the effective temperature $T_{\rm eff}$. In this way, the $T_{\rm eff}$ and $E(B{-}V)$ values are decoupled from the stellar radius.

With the mass and radius of the star now known, we can calculate $\log g$. 

Finally, the power-2 limb darkening parameters are interpolated from a grid specific to each instrument using the $\log g$ and $T_{\rm eff}$ calculated. 

The final physical variables therefore are $\sin{i}$, $\dfrac{R_{\ast}}{a}$, k, $T_{\rm eff}$, $\log g$, $\theta$, $M_{\ast}$ $\sigma_{\rm sys}$, $e$, and $\omega$.

\section{MCMC Results} \label{app:cornerplots}

\begin{figure*}
    \centering
    \includegraphics[width=17.cm, angle=0]{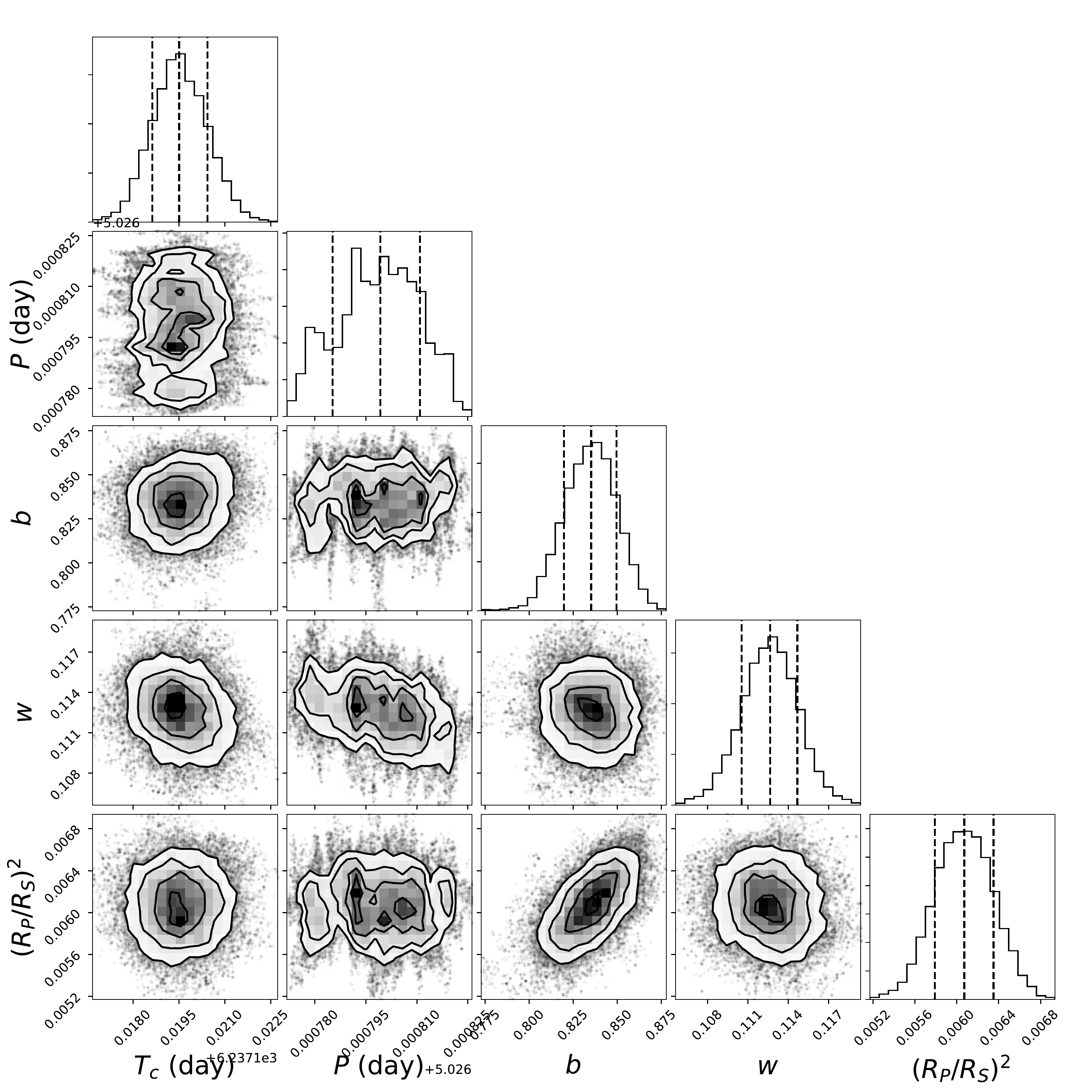}
    \caption{Cornerplot of the jump parameters describing the photometric data from the final MCMC run for WASP-186. }
    \label{fig:corner_plot_WASP186_photo}
\end{figure*}
\begin{figure*}
    \centering
    \includegraphics[width=17.cm, angle=0]{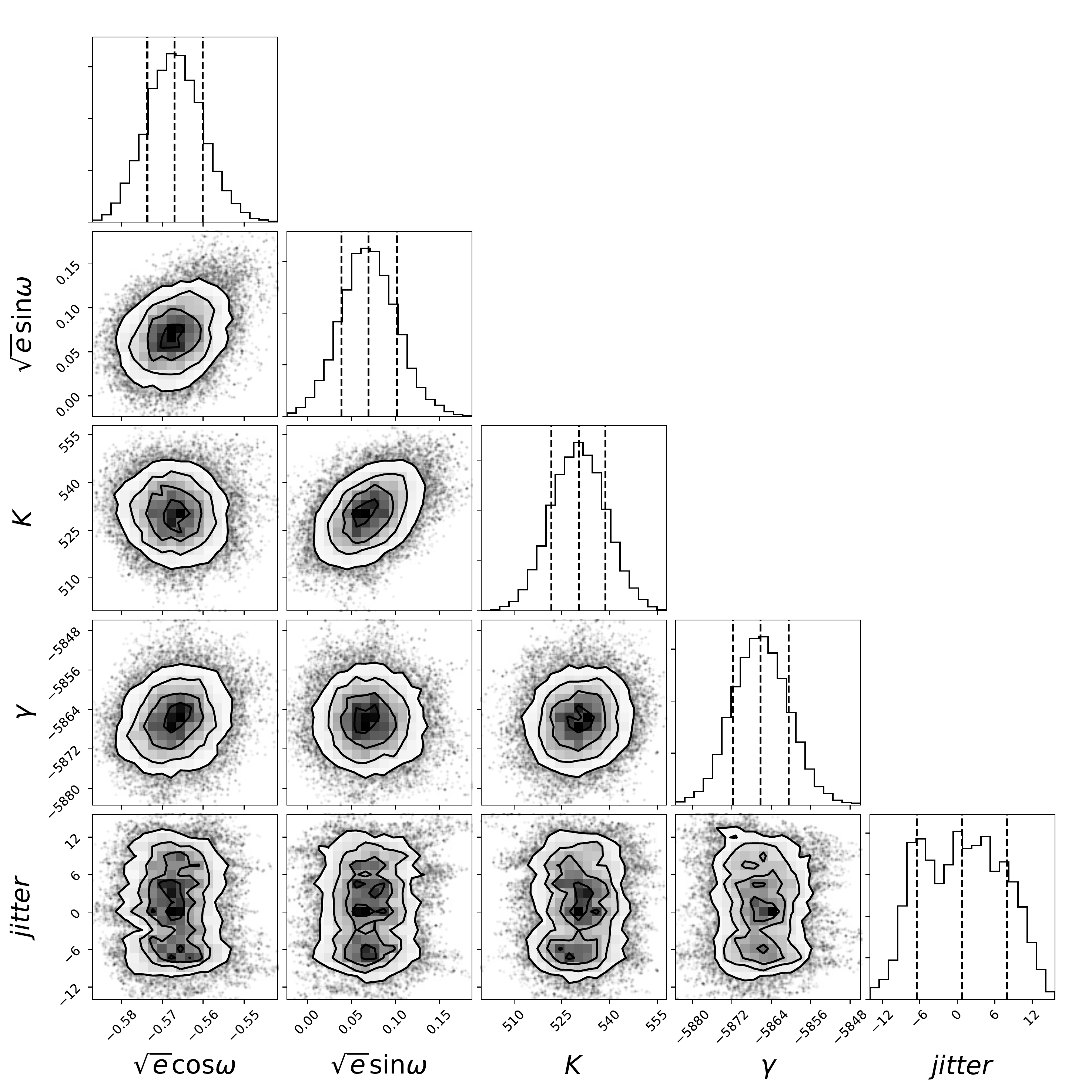}
    \caption{Cornerplot of the jump parameters describing the Radial Velocity data from the final MCMC run for WASP-186. }
    \label{fig:corner_plot_WASP186_RV}
\end{figure*}
\begin{figure*}
    \centering
    \includegraphics[width=17.cm, angle=0]{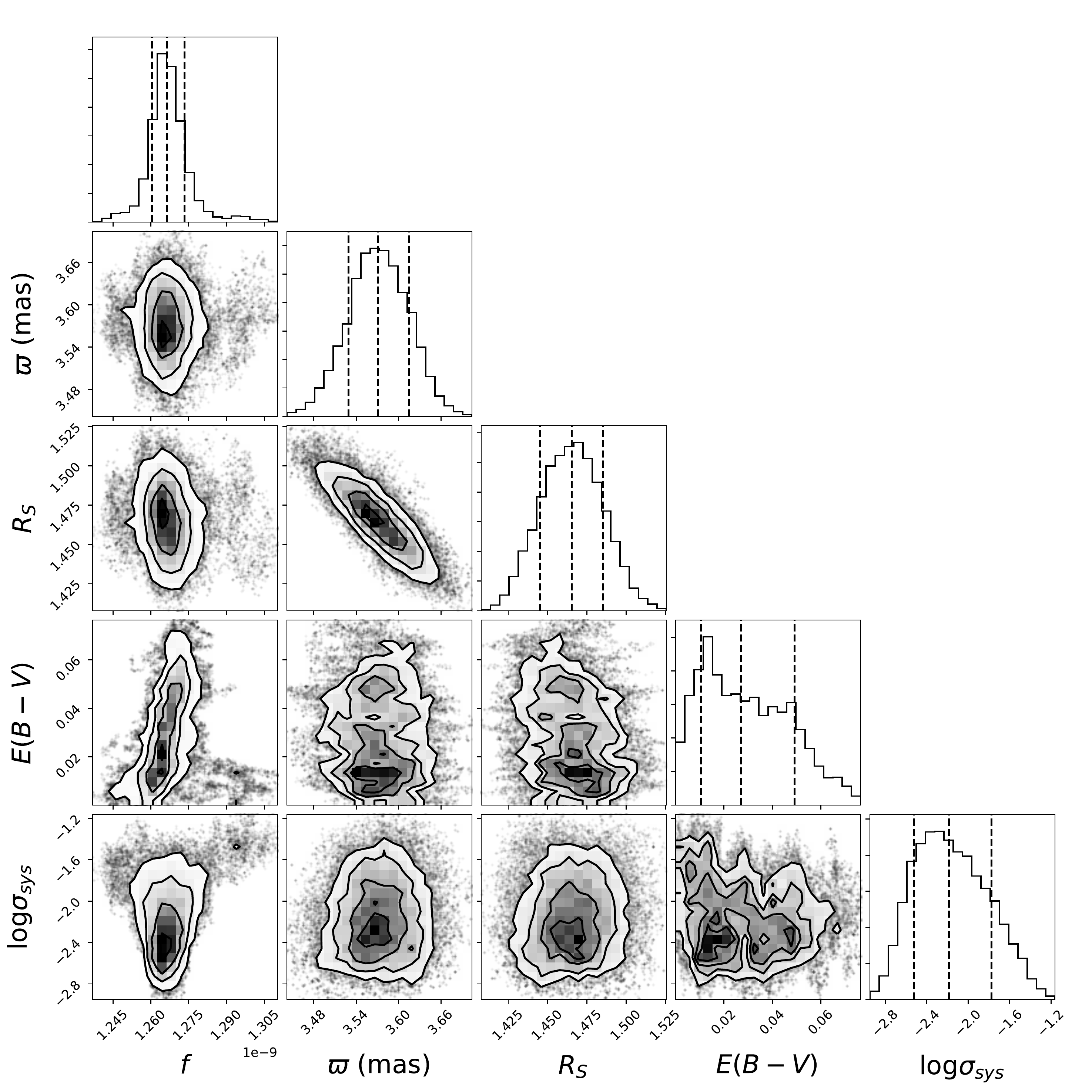}
    \caption{Cornerplot of the jump parameters describing the stellar features from the final MCMC run for WASP-186. }
    \label{fig:corner_plot_WASP186_stellar}
\end{figure*}

\begin{figure*}
    \centering
    \includegraphics[width=17.cm, angle=0]{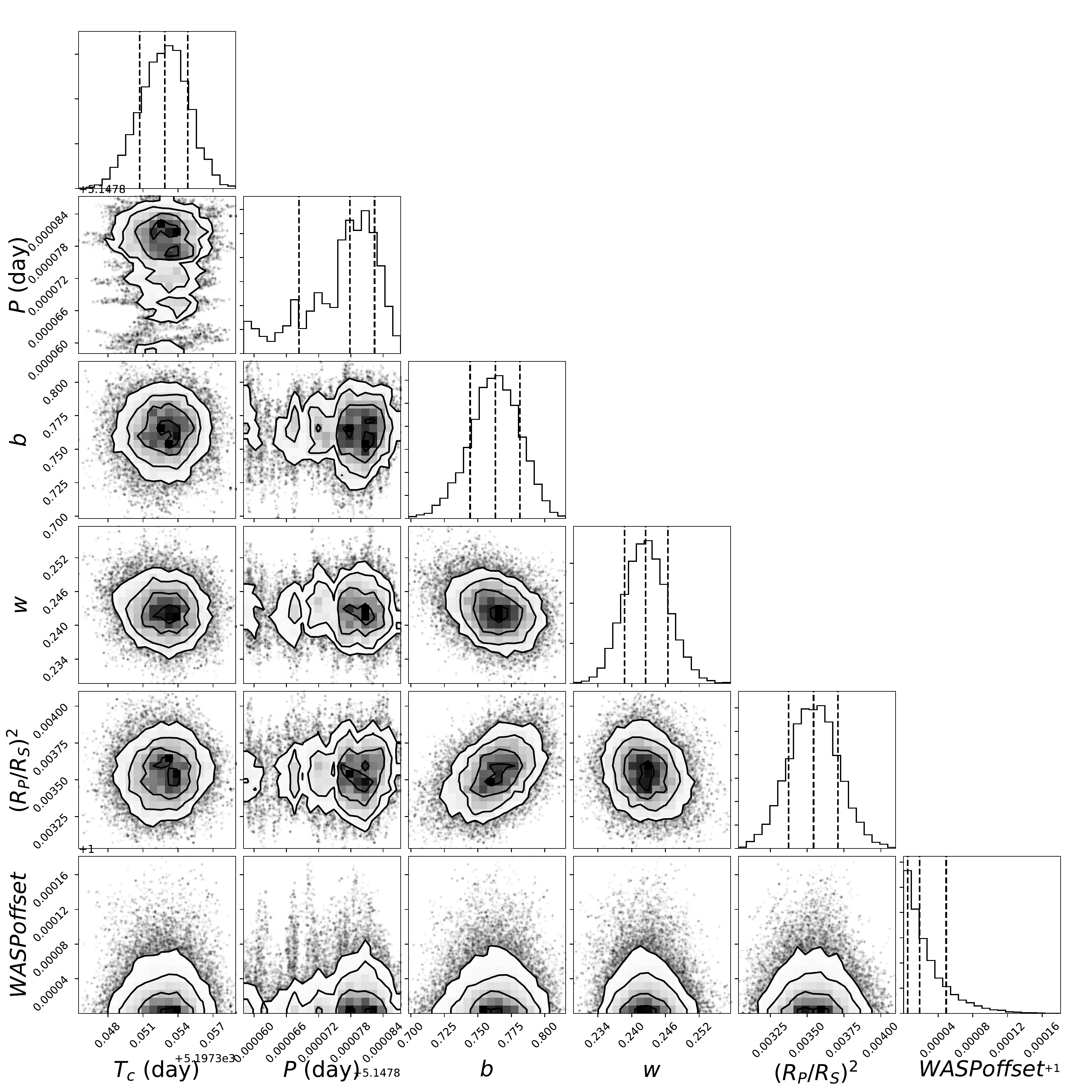}
    \caption{Same as Figure \ref{fig:corner_plot_WASP186_photo} but for WASP-187.}
    \label{fig:corner_plot_WASP187_photo}
\end{figure*}
\begin{figure*}
    \centering
    \includegraphics[width=17.cm, angle=0]{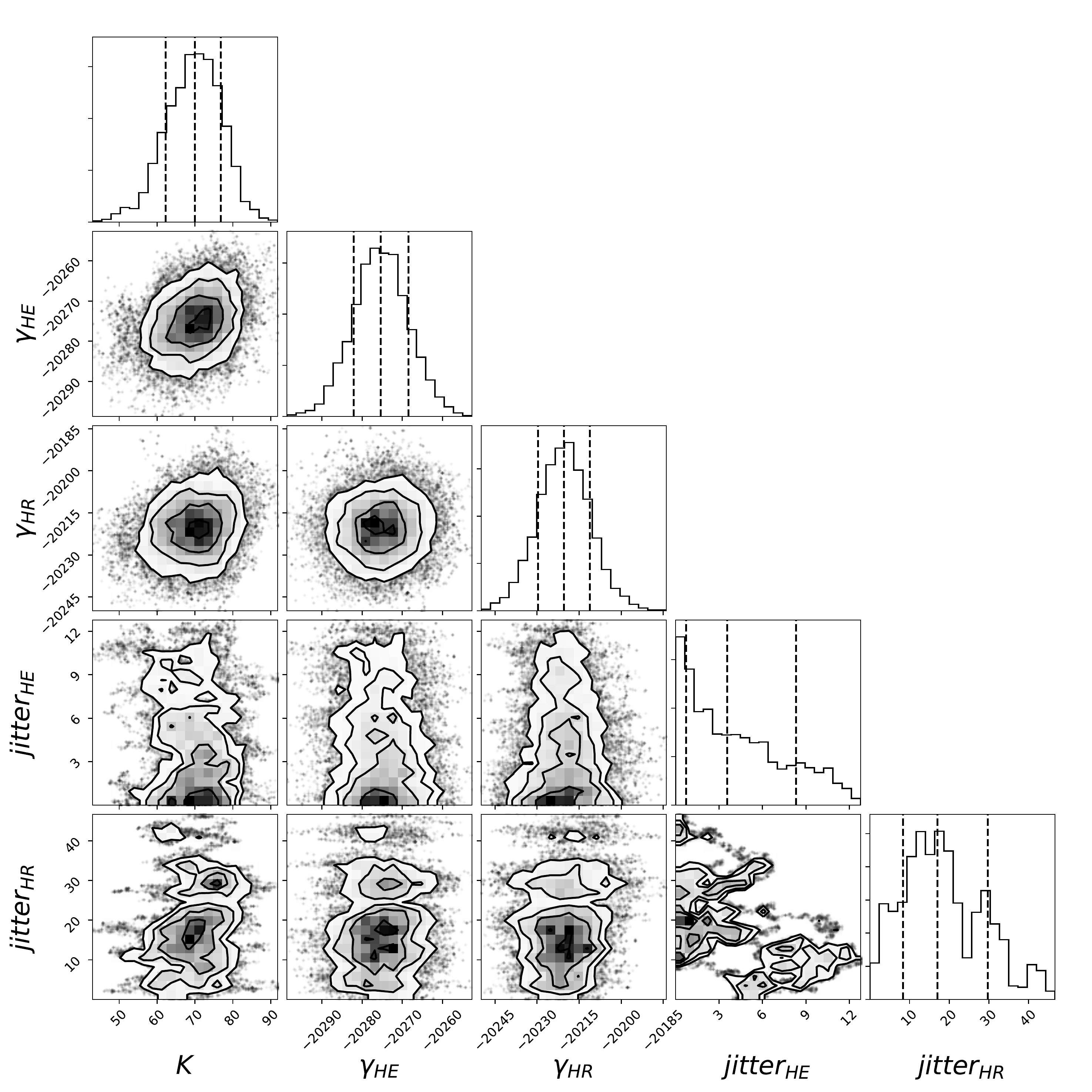}
    \caption{Same as Figure \ref{fig:corner_plot_WASP186_RV} but for WASP-187.}
    \label{fig:corner_plot_WASP187_RV}
\end{figure*}
\begin{figure*}
    \centering
    \includegraphics[width=17.cm, angle=0]{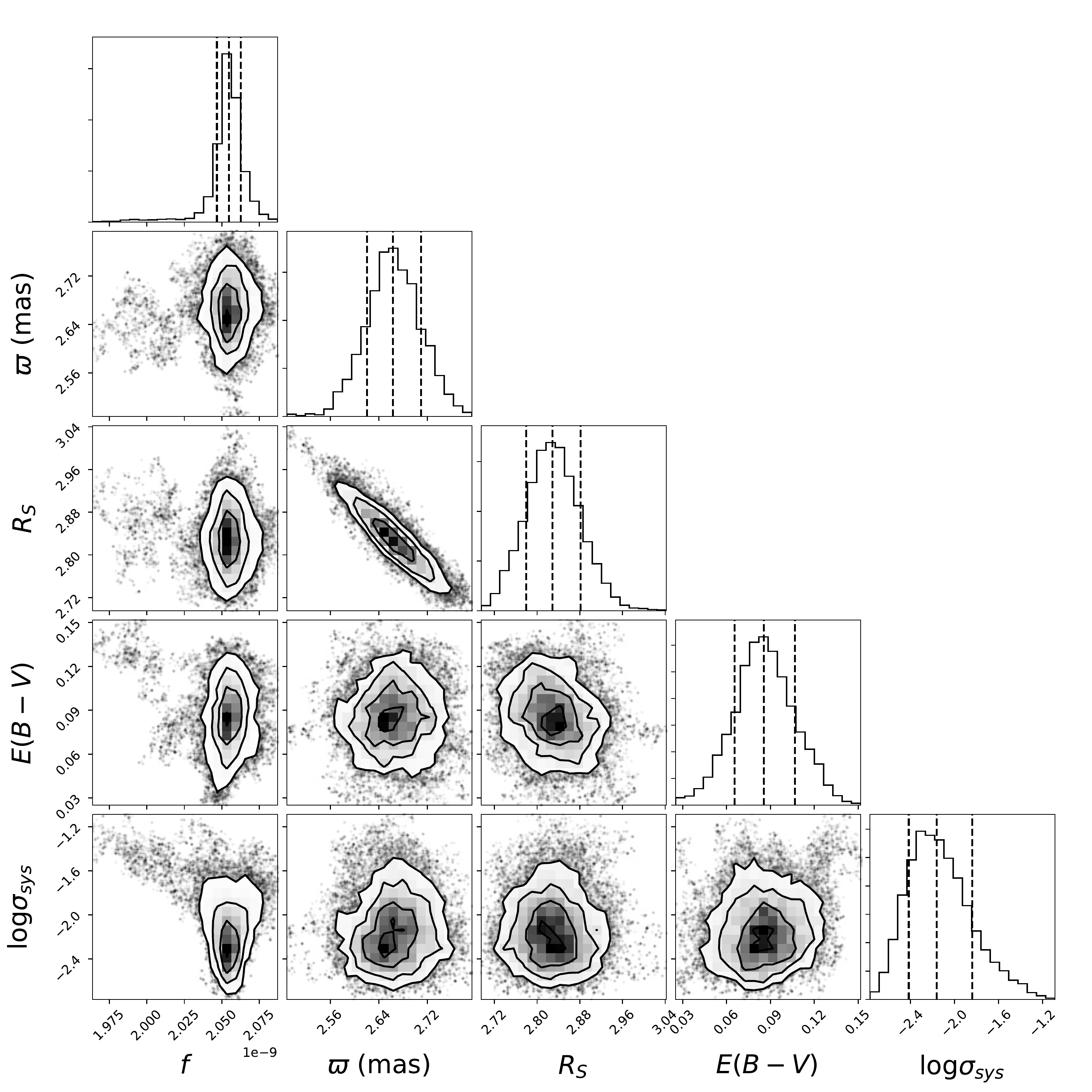}
    \caption{Same as Figure \ref{fig:corner_plot_WASP186_stellar} but for WASP-187.}
    \label{fig:corner_plot_WASP187_stellar}
\end{figure*}


\bsp	
\label{lastpage}
\end{document}